# Smartphone-based measurement of magnetic force and demonstration of Newton’s third law of motion

Sanjoy Kumar Pal[1], Soumen Sarkar[2], and Pradipta Panchadhyayee[3,4*]

[1]Anandapur H.S. School, Anandapur, PaschimMedinipur, West Bengal, India
[2]Karui P.C. High School, Hooghly, West Bengal, India
[3]Department of Physics (UG & PG), Prabhat Kumar College, Contai, PurbaMedinipur, India
[4]Institute of Astronomy Space and Earth Science, Kolkata -700054, W. B., India
[*]E-mail: ppcontai@gmail.com

**Abstract**

A fascinating approach to teaching Newton's Third Law using readily available technology is presented in this article. Magnetic forces are measured by using a smartphone's pressure sensor, two ring magnets, and common household items. Students can measure the magnitudes of forces, gain a more tangible understanding of the law, and see how 'action' and 'reaction' are quantitatively equal and opposite.

## Introduction

A novel use of a smartphone to measure the force without measuring the acceleration is discussed in this article. Demonstrating Newton's Third Law is a crucial and challenging aspect of teaching in secondary level classrooms. When students search for easily comprehensible examples to perceive the concept behind the Third Law, teachers often provide examples such as a gun's recoil after firing a bullet, jumping from a boat, bouncing a ball, etc. However, students cannot fully understand how the action and reaction become equal and opposite quantitatively, although they perceive the presence of reaction in the case of an action qualitatively. In this article, using common household items and a smartphone, we present a simple method for measurement of magnetic forces while illustrating Newton's Third Law, which leads to clear evidencethatthe reaction to an action is opposite in direction and equal in magnitude.

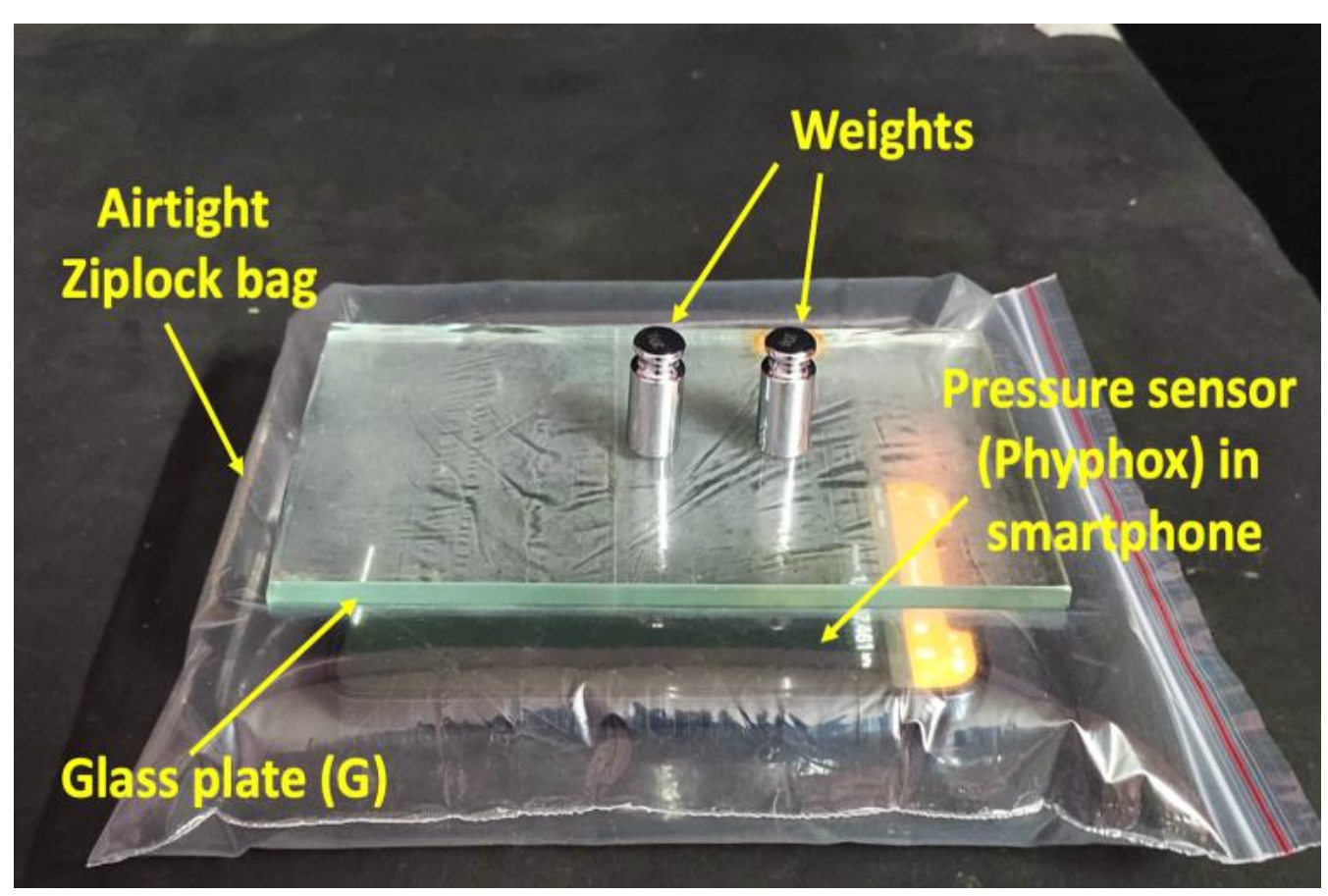

Fig. 1: Weights are placed on the glass plate (G) and the whole system is placed on an inflated Ziploc bag that encloses a Smartphone.

## Theoretical background and experiment

The use of a smartphone accelerometer sensor is reported in some articles [1] for measuring a force. Nowadays, pressure sensors are built-in features of regular smartphones. A pressure sensor provides real-time local air pressure readings via apps like Phyphox [2-4]. In this work, we have utilized a smartphone pressure sensor to measure force (weight). Initially, we turn on the pressure function on Phyphox and place the smartphone (iPhone 12 Pro Max) into an airtight Ziploc bag, ensuring enough air remains inside (see Fig. 1). To demonstrate our method, we need to determine the value of force. A hard glass plate (G) measuring (19.0 cm x 12.7 cm) is placed on the Ziploc bag. Now, we can use the pressure sensor as a weight-measuring device. The effective area upon which the force is acting differs slightly from the area of the glass plate because the Ziploc bag is not rigid. We have recorded the pressure readings for different weights on the glass plate. We plotted the values of weights versus pressure in a graph (see Fig. 2) and found the slope to be 0.2474 kg/hPa. Finally, we have obtained the effective area as 0.0242 $m^2$ (incidentally, 19 cm x 12.7 cm = 241.3 $cm^2$ = 0.02413 $m^2$) from the magnitude of the slope.

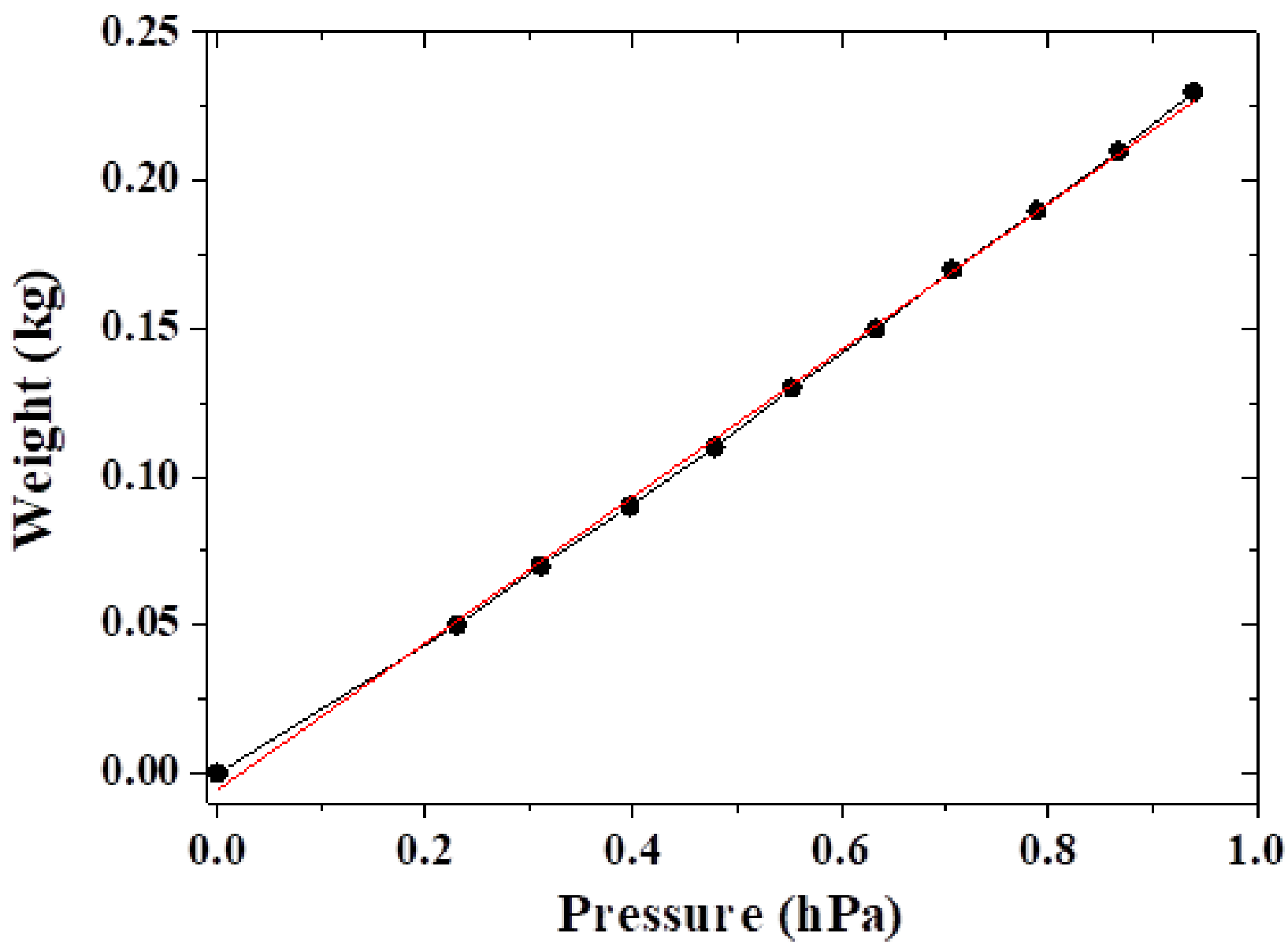

Fig. 2: Weight versus Pressure graph. The red line indicates the linear fit.

Next, a small PVC pipe is fixed vertically on the glass plate, and we take the initial pressure reading, which we denote by $p_0$. We introduce a magnet system consisting of two almost identical ring magnets (call the top magnet B, and the bottom one A). They are placed, one vertically above the other, the PVC pipe passing through the holes of the ring magnets, as shown in Fig. 3. In the following, we consider three different arrangements of the magnet system.

(I) *Use of the magnets separately*: With only the ring magnet A having a weight of $F_{G1}$, the pressure reading is, say, $p_1$ . We can write:

$$F_{G1} = (p_1 - p_0)A, \quad (1)$$

where $A$ is the effective area of the glass plate.

Similarly, instead of magnet A when the other ring magnet B having weight $F_{G2}$ is used, let the pressure reading be $p_2$. Thus, $F_{G2} = (p_2 - p_0)A$ (2)

(II) *Placing the magnets in attractive mode*: The two magnets in the magnet system are so arranged that the magnet A is placed on the glass plate and the second magnet B is placed on A in an attractive mode along the axis of the pipe. Let the magnitude of the force of B on A be represented as $F_A$ and that of A on B be $F_B$. In the attractive mode the total force on the system can be expressed as $F_1 = (F_{G1} + F_{G2} - F_A + F_B)$ taking force in the downward direction as positive. In this condition, let the pressure reading recorded be $p_3$, which obeys the following force equation:

$$F_1 = (F_{G1} + F_{G2} - F_A + F_B) = (p_3 - p_0)A \quad (3)$$

Substituting Eq. (1) and Eq. (2) in Eq. (3), we derive the following expression for $(F_B - F_A)$ :

$$(F_B - F_A) = [(p_3 + p_0) - (p_1 + p_2)]\, A. \quad (4)$$

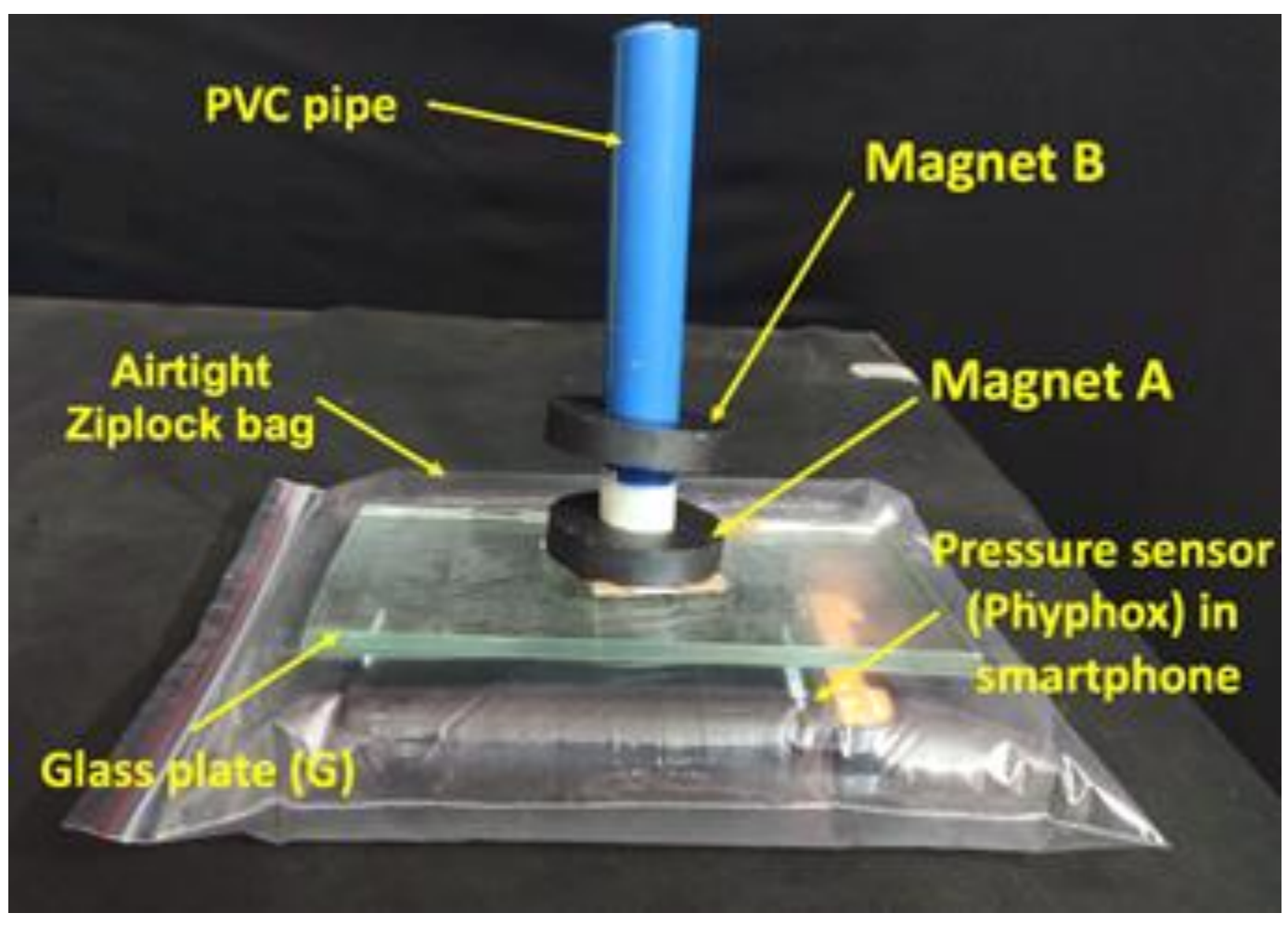

Fig. 3: The experimental setup is shown with two ring magnets mounted on the PVC pipe.

(III) *Placing the magnets in repulsive mode*: In comparison to case II, magnet B is placed on A in the repulsive mode. The magnet B is observed to float in air at a certain distance above the other magnet A. Here, the 'reaction' force $F'_A$ acts on the A magnet (lower) in the downward direction, while $F'_B$ acts in the upward direction, as a consequence of repulsion. The orientation of the magnet system in the repulsive mode results in a total downward force $F_2 = (F_{G1} + F_{G2} + F'_A - F'_B)$ as distinguished from the attractive mode.

The top magnet B seems to be weightless, leading to the increase of the apparent weight of the bottom magnet to nearly double its actual value. The equilibrium of the magnet system indicates that the weight of the top magnet B is counterbalanced by the upward repulsive force $(F'_B)$acting on B as applied by the bottom magnet A. As the weight of the B magnet is equal and opposite to the repulsive force between the two magnets:

$$F'_B = F_{G2} = (p_2 - p_0)A \quad (5)$$

By computing the pressure difference,$(p_2 - p_0)$, the repulsive force experienced by the upper magnet is estimated. Considering the total force on the magnet system we write the following expression using Eq. (5), where $P_4$ is the pressure reading noted in this orientation of the magnets:

$$F_2 = (F_{G1} + F_{G2} + F'_A - F'_B) = (F_{G1} + F'_A) = (p_4 - p_0)A \quad (6)$$

From equations (1) and (6), we obtain

$$F'_A = (p_4 - p_1)A \quad (7)$$

In Table I we have shown the pressure readings in different orientations of the magnet system. It is shown that the pressure due to two magnets in attraction mode ($p_3$) and repulsion mode ($p_4$) are nearly the same. The values of the 'action' force ($F'_B$) and the 'reaction' force ($F'_A$) are computed, which show that the average value of $F'_B$= 0.929 ± 0.002 N, which is almost equal to that of $F_A'$= 0.935 ± 0.001 N. The close agreement between the values of two counteracting forces, $F_A'$ and $F_B'$, proves that the findings support the statement of Newton's third law of motion. Again, using Eq. (4), it is clear from Table I that $F_A \approx F_B$. Thus, this result is also in conformity with Newton's third law.

**Table I: Data of the pressure values and calculation of the forces**

| *Obs. No.* | $p_0$ (hPa) | $p_1$ (hPa) | $p_2$ (hPa) | $p_3$ (hPa) | $p_4$ (hPa) | $F_A'$ (N) (Eq. 7) | $F_B'$ (N) (Eq. 5) | $(F_B - F_A)$ (N) (Eq. 4) |
|---|---|---|---|---|---|---|---|---|
| 1 | 1011.730 | 1012.115 | 1012.115 | 1012.499 | 1012.500 | 0.933 | 0.933 | 0.0000242 |
| 2 | 1011.729 | 1012.113 | 1012.114 | 1012.500 | 1012.499 | 0.936 | 0.933 | 0.0000484 |
| 3 | 1011.732 | 1012.116 | 1012.114 | 1012.502 | 1012.501 | 0.933 | 0.926 | 0.0000968 |
| 4 | 1011.733 | 1012.114 | 1012.115 | 1012.502 | 1012.502 | 0.941 | 0.926 | 0.0001452 |
| 5 | 1011.730 | 1012.114 | 1012.113 | 1012.500 | 1012.499 | 0.933 | 0.928 | 0.0000726 |

In Eqs. (3) and (6), if the signs of $F_A$ and $F_B$ (or of $F_A'$ and $F_B'$) are not put in by hand, the consideration of the experimental constraint of equality of the forces $F_1$ and $F_2$ given by the expressions $F_1 = (F_{G1} + F_{G2} + F_A + F_B)$ and $F_2 = (F_{G1} + F_{G2} + F_A' + F_B')$, respectively, imply that $F_A$ (or $F_A'$) is equal and opposite to $F_B$ (or $F_B'$).

**Acknowledgement**:
We gratefully acknowledge Dr Debapriyo Syam for stimulating discussion and his contribution to the final preparation of the manuscript.